\documentclass{emulateapj}
\usepackage{apjfonts}
\usepackage{lscape}

\newcommand{\solarmass}{\mbox{${\rm M_{\odot}}$}}

\def\ltsima{$\; \buildrel < \over \sim \;$}
\def\simlt{\lower.5ex\hbox{\ltsima}}
\def\gtsima{$\; \buildrel > \over \sim \;$}
\def\simgt{\lower.5ex\hbox{\gtsima}}

\newcommand{\grs}{{GRS 1915+105}}

\newcommand{\etal}{{et al.}}

\newcommand{\rxte}{{\it RXTE}}
\newcommand{\cgro}{{\it CGRO}}

\newcommand{\chandra}{{\it Chandra}}

\newcommand{\suzaku}{{\it Suzaku}}

\lefthead{Ueda et al.}
\righthead{Suzaku Observation of \grs }

\begin{document}

\title{\suzaku\ Observation of GRS 1915+105:\\
Evolution of Accretion Disc Structure during Limit-Cycle Oscillation}

\author{Y.~Ueda\altaffilmark{1}, K.~Honda\altaffilmark{2}, H.~Takahashi\altaffilmark{2}, C.~Done\altaffilmark{3}, \\
H.~Shirai\altaffilmark{2}, Y.~Fukazawa\altaffilmark{2}, K.~Yamaoka\altaffilmark{4}, S.~Naik\altaffilmark{5}, \\
H. Awaki\altaffilmark{6}, 
K. Ebisawa\altaffilmark{7}, 
J. Rodriguez\altaffilmark{8}, 
S. Chaty\altaffilmark{8}}

\altaffiltext{1}{Department of Astronomy, Kyoto University, Kyoto 606-8502, Japan}
\altaffiltext{2}{Department of Physical Science, Hiroshima University, Hiroshima739-8526, Japan}
\altaffiltext{3}{Department of Physics, University of Durham, South Road, Durham DH1 3LE, UK}
\altaffiltext{4}{Department of Physics, Aoyama Gakuin University, Sagamihara, Kanagawa 229-8558, Japan}
\altaffiltext{5}{Astronomy \& Astrophysics Division, Thaltej Campus, Physical Research Laboratory, Ahmedabad, India}
\altaffiltext{6}{Department of Physics, Faculty of Science, Ehime University, Matsuyama 790-8577, Japan}
\altaffiltext{7}{Institute of Space and Astronautical Science/JAXA, Sagamihara, 229-8510, Japan}
\altaffiltext{8}{Laboratoire AIM, CEA/DSM-CNRS-Universit\'e Paris Diderot, 
Irfu/Service d\'Astrophysique, B\^at. 709, CEA-Saclay, 91191 Gif-sur-Yvette Cedex, France}

\begin{abstract}

We present results from the \suzaku\ observation of the microquasar
\grs\ performed during the 2005 October multiwavelength campaign. The
data include both stable state (class $\chi$) and limit-cycle
oscillation (class $\theta$). Correct interstellar absorption as well
as effects of dust scattering are fully taken into account in the
spectral analysis. The energy spectra in the 2--120 keV band in both
states are all dominated by strong Comptonization of disk photons by
an optically thick ($\tau \approx$7--10) and low temperature
($T_{\rm e} \approx$2--3 keV) hybrid plasmas containing
non-thermal electrons produced with 10--60\% of the total power
input. Absorption lines of highly ionized Fe ions detected during the
oscillation indicate that a strong disk wind is developed. The
ionization stage of the wind correlates with the X-ray flux,
supporting the photoionization origin. The iron-K emission line shows
a strong variability during the oscillation; the reflection is
strongest during the dip but disappears during the flare. We interpret
this as evidence for ``self-shielding'' that the Comptonizing corona
becomes geometrically thick in the flare phase, preventing photons
from irradiating the outer disk. The low-temperature and high
luminosity disk emission suggests that the disk structure is similar
to that in the very high state of canonical black hole binaries. The
spectral variability during the oscillation is explained by the change
of the disk geometry and of the physical parameters of Comptonizing
corona, particularly the fractional power supplied to the acceleration
of non-thermal particles.

\end{abstract}

\keywords{accretion, accretion disks --- stars: individual (\grs ) ---
        techniques: spectroscopic --- X-rays: stars}

\section{Introduction}

\grs\ is the brightest microquasar in our Galaxy, providing us with a
unique opportunity to study the accretion flow onto a black hole at
high fractions of Eddington ratio \citep[for a review,
see][]{fen04}. This source, discovered in 1992 with the WATCH
instrument on GRANAT \citep{cas92}, has been persistently active up to
now unlike usual soft X-ray transients. \grs\ was recognized to be a
superluminal source by VLBA observations of radio outbursts
\citep{mir94}, making it a target of great importance for studying the
formation mechanism of relativistic jets. From the kinetics of the
radio jets, their intrinsic speed and inclination are determined to be
$0.92 c -0.98 c$ ($c$ is the light speed) and $66^\circ-70^\circ$,
respectively \citep{mir94,fen99}. The near infrared observations have
revealed that the system consists of a black hole with a mass of
$14\pm4$ \solarmass\ and a K~III type companion star with an orbital
period of 33.5 days \citep{gre01}.  This indicates a huge size of the
accretion disk with a continuously high mass transfer rate from the
companion, which may account for many unique features of this source
compared with normal black hole binary systems.

The source has been intensively studied at high energies by many
observatories on different occasions. \rxte /PCA observations revealed
that \grs\ often exhibits dramatic temporal and spectral variations,
so-called limit-cycle oscillations, unique features in accreting
stellar mass black holes in our Galaxy
\citep[e.g.,][]{bel97}. \citet{bel00} found the instantaneous spectral
state of \grs\ can be divided into states, States~A, B, and C, where
the source undergoes frequent transition or stays stable. Based on
these transition patterns, they classified them into 12 Classes,
including 2 ``stable'' classes (Class~$\chi$ and $\phi$) with little
variability on a time scale longer than $\sim 1$ sec. This behavior is
quite different from normal black holes observed in canonical states,
the low/hard, intermediate (or very high), and high/soft states
(\citet{tan95}; see also \citet{hom05} and \citet{rem06} for more
recent classification). The correspondence of States~A, B, and C of
\grs\ to these states is not completely understood, however
\citep[][]{rei03}.

Theoretically, the limit-cycle oscillation can be explained by thermal
instability in the inner part of the accretion disk under high mass
accretion rate, triggering limit-cycle transitions between two stable
branches in the surface density vs $\dot{M}$ plane: the standard disk
and optically-thick advection dominated accretion flow, so-called the
slim disk \citep{abr88}. In fact, two-dimensional hydrodynamic
simulations are successful to reproduce the limit cycle behaviors of
\grs\ qualitatively \citep{ohs06}. However, direct observations that
trace the evolution of the disk structure both in the oscillation and
stable state are still insufficient to be compared with model
predictions, and their interpretations may be subject to large
uncertainties since understanding of the origin of the X-ray emission
is not fully established. Many early works relied on the spectral
modelling by a multi-color disk (MCD; \citealt{mit84}) model plus a
power law, as applied to other black hole binaries. There is no
guarantee that this model always holds to \grs, in particular when the
power law component dominates the entire flux.

Later observational works \citep{zdz01,don04,ued09} have suggested that
the X-ray spectra of \grs\ at least in some states are dominated by an
optically-thick Comptonization of the disk photons off low energy
electrons, by applying physically more realistic models than the
canonical MCD + power law model. \citet{zdz01,zdz05}
analyzed broad band spectra observed with \rxte /PCA, HEXTE, and \cgro
/OSSE, and found that they can be described by Comptonization from
thermal/non-thermal hybrid plasmas. Such low temperature,
optically-thick Comptonization may be common characteristics in
accretion flows at high Eddington fractions, which may apply to those
onto supermassive black holes \citep{mid09}.

Observations of local spectral features such as iron-K emission line
and absorption lines give key information to reveal the structure of
accretion disks. From high resolution spectra (CCD or the HETGS
instrument on \chandra ) of \grs, absorption lines from highly ionized
ions have been clearly detected \citep{kot00,lee02,ued09}. This
indicates the presence of strong disk wind most probably occurring at
$r\sim 10^5 r_{\rm g}$ ($r_{\rm g} \equiv GM/c^2$ is the gravitational
radius where $G$, $M$, and $c$ being the gravitational constant, mass
of the black hole, and light velocity, respectively), which carries a
huge amount of accreting gas outward the system
\citep{ued09,nei09}. On the other hand, the emission line profile
gives constraints on the geometry between the continuum emitting
region and reflector, most probably the accretion disk. An iron-K
emission line has been detected from \grs, whose shape and intensity
seem to depend on the states \citep[][]{kot00,nei09}. Since its
equivalent width is not large ($<50$ eV) in \grs, we need both good
spectral resolution and large effective area to accurately measure the
profile, in particular to trace the change during oscillations.

\suzaku\ \citep{mit07}, the 5th Japanese X-ray satellite, observed
\grs\ on 2005 October 16--18 (UT throughout the paper) as a part of
the science working group's observation program. The large effective
area with good energy resolution in the 0.2--12 keV band, together
with the simultaneous coverage of the 10--600 keV band with
unprecedented sensitivities, give us the best opportunities to reveal
the origins of the X-ray emission and disk structure of \grs\ both in
stable and oscillation states. In our work, we refer to the latest
results on the metal abundances of interstellar (plus circumstellar)
gas toward \grs\ as determined by \chandra /HETGS \citep{ued09}. In
the analysis, we also take into account the effects by
dust-scattering, which have been ignored in most of previous works. We
show that these are important for accurate modelling of the
spectra. \S~2 describes the observation and data reduction, and \S~3
the light curves and states of \grs\ in our observations. The spectral
analysis is presented in \S~4. We discuss the interpretation of our
results in \S~5. The conclusions are summarized in \S~6. In our paper,
we assume the distance of $D=12.5$ kpc and inclination angle of
$i=70^\circ$ unless otherwise stated. Around the epoch of our \suzaku\
observations, a large multiwavelength campaign was conducted involving
other space and ground observatories, whose results will be reported
in a separate paper (Ueda et al., in preparation). For the preliminary
results of the campaign, refer to \citet{ued06}. The {\it INTEGRAL}
light curves are published in \citet{rod08a}.

\section{Observations and Data Reduction}

\suzaku\ observed \grs\ from 2005 Oct 16 16:42 to Oct 18 23:16 for a
net exposure of $\approx$80 ksec. \suzaku\ carries four sets of X-ray
Telescope (XRT) each with a focal plane X-ray CCD camera, the X-ray
Imaging Spectrometer (Front-side Illuminated XISs, XIS-0, XIS-2, and
XIS-3; Back-side Illuminated XIS, XIS-1), and a non-imaging collimated
instrument called the Hard X-ray Detector (HXD), which consists of the
silicon $p$-intrinsic-$n$ photo-diodes (PIN) and scintillation
counters made of gadolinium silicate crystals (Ce-doped Gd$_2$SiO$_5$;
GSO). The XIS, PIN, and GSO simultaneously covers the energy band of
0.2--12 keV, 10--60 keV, and 40--600 keV, respectively. The target was
observed at the so-called XIS-nominal position.

The XISs were operated in the normal clock mode with the $2\times2$
editing mode for XIS-0, 2, and 3, and $3\times3$ for XIS-1. To
minimize pile-up (double photon events) due to the brightness of this
source, the 1/8 window option was adopted for XIS-0, XIS-2, and XIS-3,
which reduced the exposure time per frame from 8 sec to 1 sec. XIS-1
was operated with the 0.1 sec burst plus 1/8 window option, although
the 1/8 window option was turned off for XIS-1 before Oct 18 00:26,
due to an accidental error in the satellite operation. We utilize XIS
data taken with different modes according to the purpose of an
analysis. The XIS-1 events are used to derive (1) unsaturated light
curves with a resolution of 1 sec (after Oct 18 00:26) and 8 sec
(before then) and (2) continuum spectra with little effects from
pile-up, while we use XIS-0, 2 and 3 data to make the spectra with
high photon statistics in the 5--10 keV band for detailed study of the
iron-K features.

We analyze the data using the HEAsoft version 6.4.1 package, utilizing
the products version 2.0.6.13 processed by the pipeline processing
team and calibration database (CALDB) released on 2008 Dec.\ 3. For
the analysis of the XIS data, we exclude events suffering from
telemetry saturation, based on the GTI filters provided by the XIS
team\footnote{http://www-cr.scphys.kyoto-u.ac.jp/menber/hiroya/xis/gtifile.html}.
We do not subtract the non X-ray background nor the cosmic X-ray
background, which are negligible compared with the bright source
counts. For the spectral analysis of 
the XIS-1 data taken with the full window mode before Oct 18
00:26, we subtract out-of-time events as background by utilizing
events in an outer region in the same CCD chip. In the extraction of
the spectra of XIS-0, 2, and 3 (i.e., without burst options), we
exclude a circular region with a radius of 120 pixels around the
target position to reduce the effects of pile-up events. It is known
that the attitude of the \suzaku\ satellite exhibits an orbital
variation by $\sim$0.5 arcmin, causing periodic shifts of the target
position in the focal plane. Hence, we check the central position of
the source and define an event extraction region every 200 seconds.

For the HXD, we use the best modelled background files
provided by the HXD 
team\footnote{http://www.astro.isas.ac.jp/suzaku/analysis/hxd/pinnxb/tuned/
(PIN) and http://www.astro.isas.ac.jp/suzaku/analysis/hxd/gsonxb/
(GSO)}. The reproductivity of the NXB spectra with a 10 ksec exposure
is known to be 1.4\% and 0.6\% for the PIN (15--40 keV) and GSO
(50--100 keV), respectively \citep{fuk09}. Then, an estimated CXB
spectrum is added to the NXB spectrum of the PIN with the angular
transmission function for a uniformly extended emission. The CXB is
neglected in analysis of the GSO data.

\section{Light Curves and X-ray States}

Figure~1 shows the light curves of \grs\ obtained with XIS-1 in the
1--10 keV band and with the HXD/PIN in the 10--60 keV during the whole
observation epoch. All photons observed in the XIS-1 chip are utilized
here, but the XIS-1 count rates before Oct 18 00:26 are artificially
reduced by a factor of 0.51 to correct for the difference of the
integration area from the data taken with the 1/8 window option.  For
convenience, hereafter we define $t$ as the time since 2005 October
16.0 in units of hours. It is seen that at $t\simlt40$ \grs\ was in an
X-ray stable state, except for a possible flare-like event seen at
$t\simeq21.5$. From $t\simeq40$, the X-ray flux increased rapidly and
entered into ``oscillation'' that lasted till $t\simeq56$. From
$t\simeq66$, \grs\ exhibits a similar X-ray flare, followed by
oscillation whose variability pattern is different from the previous
one (see \citealt{ued06} and \citealt{rod08a} for X-ray light curves
observed with other satellites). Throughout the observation, the
source was very bright in soft X rays, $\sim (1-3)\times 10^{-8}$ erg
cm$^{-2}$ s$^{-1}$ in the 2--10 keV band.

\begin{figure}
\epsscale{1.2}
\plotone{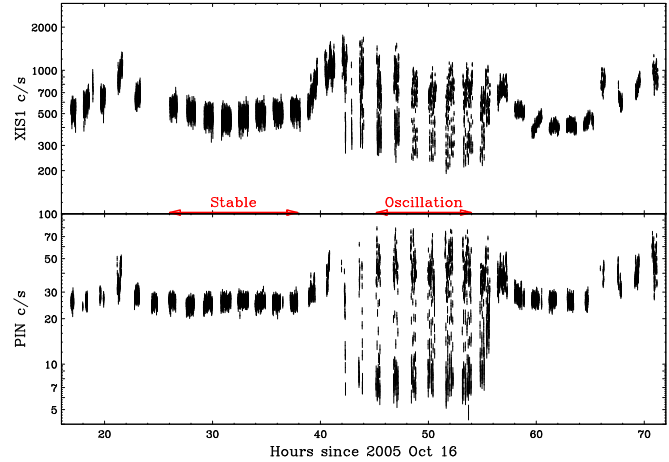}
\caption{
The light curves of \grs\ in the 1--10 keV and 10--60 keV band
obtained with XIS-1 and HXD/PIN, respectively.
The data are binned in intervals of 16 seconds. 
The red arrows denote the time regions used for the spectral
analysis of the stable and oscillation states.
}
\label{fig1}
\end{figure}

For comparison with previous studies of \grs\, we first identify the
``Class'' and ``State'' of \grs\ in our observations, according to the
model-independent classification scheme by \cite{bel00}. As seen from
Figure~1, \grs\ continuously stayed in the ``stable state'' in the
period of $t=26-38$, which is similar to the plateau
state~\citep{fos96}, before entering into the oscillation
phase. Timing and spectral analyses of the data taken with the
proportional counter arrays (PCA) on \rxte\ at the epoch of $t=23-26$
reveal that our ``stable state'' corresponds to Class~$\chi$ (stable
State~C) having an extremely soft spectrum, accompanied with a quasi
periodic oscillation (QPO) at $\approx$6 Hz \citep{ued06}.

Figure~2 shows representative X-ray light curves ($t=51.6-52.3$) in
the oscillation phase taken with the \suzaku\ XIS-1 and PIN in the
1--10 keV and 10--60 keV band, respectively, together with their
hardness ratio. We identify the X-ray variability pattern as
Class~$\theta$ in \citet{bel00}, where the transition occurs between
State~C and State~A (soft dip) on a time scale of $\sim$10 minutes. It
is seen that during oscillations the spectral colors significantly
softens in the dip phase, as expected according to previous studies
with \rxte\ \citep{bel00}.

\begin{figure}
\epsscale{1.2}
\plotone{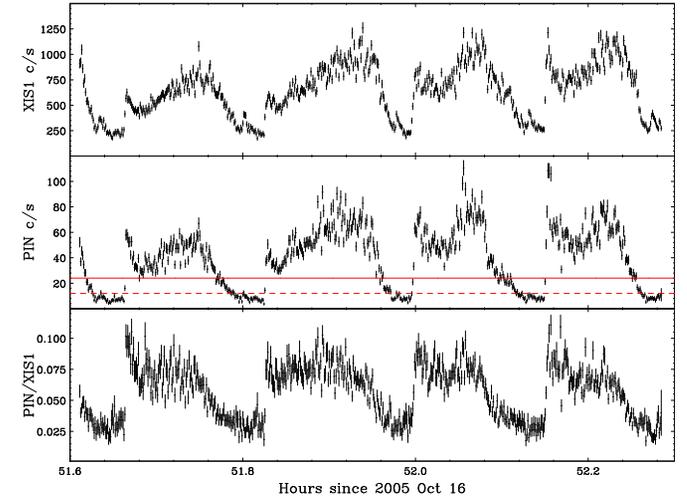}
\caption{A blow up of XIS-1 (1--10 keV) and PIN (10--60 keV) light
  curves at $t=51.6-52.3$ and their hardness ratio with 4 sec bin.
The two horizontal lines
  indicate the limits between the oscillation low/med and med/high
  states, respectively.
}
\label{fig2}
\end{figure}

\section{Spectral Analysis}

\subsection{Overview}

For spectral analysis, we define four different states (a) ``stable
state'' (Stable), (b) ``oscillation high state'' (Osc-H),
(c)``oscillation medium state'' (Osc-M), and (d)``oscillation low
state'' (Osc-L), according to time variability pattern and intensity
level. The stable and oscillation-high states consist of State~C data
defined by \citet{bel00}, while oscillation-med and low corresponds to
State~A. The time region is taken between $t=26.03-37.88$ for (a),
when the source flux was almost constant (Class~$\chi$), and
$t=45.18-53.97$ for the rest (Class~$\theta$). The oscillation high,
med, low states are defined as when the 10--60 keV PIN count rate is
above 24 c s$^{-1}$, between 12--24 c s$^{-1}$, and between 1--12 c
s$^{-1}$, respectively, within the above time region. The count rate
thresholds for the oscillation states are illustrated by two lines in
the middle panel of Figure~2. Table~1 summarizes the definition of the
states. We make final good time intervals (GTIs) by combining the XIS,
PIN, and GSO, to cover exactly simultaneous time regions in each
state. The net exposures are 19.8 ks, 7.2 ks, 0.93 ks, and 2.6 ks for
the stable, oscillation-high, med, and low state, respectively.

\tabcolsep=2pt
\begin{deluxetable}{ccccc}
\tablenum{1}
\tablecaption{State Definition\label{tbl-1}}
   \tablehead{\colhead{State} & \colhead{Start} & \colhead{End} &
     \colhead{Criteria} &\colhead{Exposure} \nl
     \colhead{}& \colhead{(UT)}& \colhead{(UT)}&
     \colhead{PIN (10--60 keV)}& \colhead{(ksec)}
   }
\startdata
Stable & 2005 Oct. 17 02:02& Oct. 17 13:53& \nodata & 19.8\nl
Osc-H & 2005 Oct. 17 21:11& Oct. 18 05:58& $>$24 c/s& 7.2\nl
Osc-M & 2005 Oct. 17 21:11& Oct. 18 05:58& 12--24 c/s& 0.93\nl
Osc-L & 2005 Oct. 17 21:11& Oct. 18 05:58& 1--12  c/s& 2.6\nl
\enddata
\end{deluxetable}

The spectral analysis we perform consists of two steps, (1) analysis
of local features in the 5--10 keV band, which determines the
parameters of a reflection component (see below), and (2) that of
continuum emission over the 2--120 (or 2--50) keV band. As mentioned
above, different XIS sensors are used for each purpose. For the former
step, we use XIS-0, 2, and 3 (FI-XISs), which were operated without
the burst options, by excluding the central core of the point spread
functions; this makes us achieve good photon statistics by avoiding
extreme pile-up effects. The spectra of the three sensors are summed
together with the energy responses. For the latter step, we make a
simultaneous spectral fit to the spectra of XIS-1 (BI-XIS), PIN, and
GSO, covering the 2--120 keV band, for the stable and oscillation-high
states, while we only use the XIS and PIN spectra in the 2--50 keV
band for the oscillation-med and low states, since the GSO data in
these states have limited statistics due to the low flux and short
exposure. The data of XIS-1, operated with the 0.1 s burst option, is
almost free from pile-up effects and we can utilize all photons in the
point spread function. Thanks to this fact, we can best estimate the
continuum below 10 keV with least uncertainties in the calibration of
the energy response of the XIS+XRT system, although the photon
statistics is poorer than the data of the three FI-XISs. The observed
spectra of the XIS-1 and HXD in the stable and oscillation state are
plotted in Figure~3 (a) and (b), respectively.

\begin{figure*}
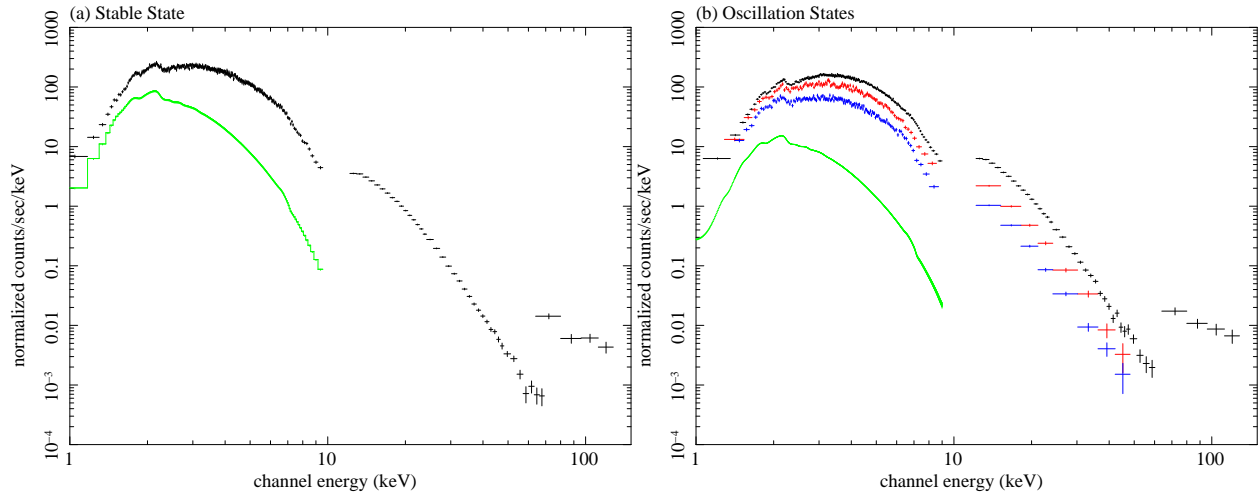

\epsscale{1.0}
\begin{center}
\includegraphics[angle=270,scale=0.35]{f3a.ps}
\includegraphics[angle=270,scale=0.35]{f3b.ps}
\end{center}
\caption{
(a) ({\it left}) The observed spectra of XIS-1 (full window mode) and
the HXD in the stable state, folded by the detector response. (b)
({\it right}) Those of XIS-1 (1/8 window mode) and the HXD in the
oscillation-high (upper, black including the GSO), oscillation-med
(middle, red), and oscillation-low (lower, blue) states.  The lower
curves in the XIS-1 band (green) indicate the estimated contribution
of the dust scattered component. }
\label{fig3}
\end{figure*}

Since the determination of the local features affects that of the
continuum and vice versa, the above two steps are repeated in an
iterative way. Firstly, we roughly determine the continuum shape from
the BI-XIS and HXD spectra. Secondly, using the FI-XISs spectra in the
5--10 keV band, we determine local iron-K features, including
absorption lines and an iron-K emission line, which is modelled by a
feature from a reflection component. In this stage, we fix the
parameters of the continuum at the best-fit obtained above except for
its normalization. To take into account a remaining component of
pile-up events, we add a power law component having a positive photon
index in the model, whose parameters are set free. Thirdly, we again
perform the continuum fit of the BI-XIS and HXD spectra, but including
the reflection component and absorption line features, whose
parameters are fixed at the best-fit values as determined in the
previous fit of the FI-XISs spectra. Detailed description of the
continuum model and local features is given in the following
subsections.

\subsection{Absorption and Dust Scattering}

In our study, special care is paid on two important effects that could
affect the modelling of the continuum, (1) interstellar absorption and
(2) dust scattering. It is well known that the elemental abundances in
interstellar medium in our Galaxy are different by location, and are
not ``Solar'', as often assumed in spectral analyses of Galactic
sources. To model the total photo-electric absorption toward \grs, we
refer to the results by \citet{ued09} obtained from the \chandra
/HETGS spectrum, in terms of ``equivalent'' hydrogen column densities
(converted from the Solar abundances by \citealp{and89}) of
$2.78\times10^{22}$ cm$^{-2}$ for H, He, C, N, and O,
$6.3\times10^{22}$ cm$^{-2}$ for Ne, Na, Mg, and Al,
$5.8\times10^{22}$ cm$^{-2}$ for Si, $7.6\times10^{22}$ cm$^{-2}$ for
S, Cl, Ar, and Ca, and $10\times10^{22}$ cm$^{-2}$ for Cr, Fe, Co, and
Ni. The absorption cross section by \citet{wil00} is adopted, which is
available as {\it TBvarabs} model in XSPEC.

We also take into account the effects by dust scattering, which are
more significant in soft X-rays particularly below $\sim 3$ keV. A
part of direct emission from a heavy absorbed source is scattered out
by interstellar dust in the line-of-sight, while photons emitted with
slightly different angles are scattered in, thus making
dust-scattering halo around the target. At first order approximation
the scattered-in and scattered-out lights cancel each other, although
there is time delay before scattered-in components reach an observer
following the direct component. Hence, we do need to correct for this
effect as far as the source spectrum is constant {\it and} one can
integrate all the emission including both direct and halo components.

This is not the case for \grs\ observed with \suzaku, however, because
(1) the halo components cannot be fully collected in a limited field
of view of the XIS and (2) the source is highly variable in the
oscillation states. Assuming the same image profile of the dust
scattered halo as GX 13+1 \citep{smi02}, which has a similar
absorption column density to \grs, we find that the effective area of
the XIS-1 (full window) for the halo component is $\approx$60\% of
that for the direct (i.e., point source) one, and thus the scattering
effect is not canceled out even when the spectrum is constant. When it
is variable, the situation becomes more complex; the total spectrum
observed in a given time interval is contaminated by halo components
emitted in earlier epochs due to their time delay. The time scale of
the delay is estimated to be $\sim \theta^2 D/c$, where $\theta$ and
$D$ is the halo size and distance to the source, respectively. Taking
$\theta \sim 2'$ according to the halo profile of GX 13+1, we estimate
the delay to be $\sim$100 hours, much longer than the time scale of
the oscillations. This means that we approximately observe the
spectrum of {\it variable} direct emission plus {\it constant}
dust-scattered component in the oscillation states. According to the
light curve of all sky monitor on \rxte\ \citep{ued06}, the averaged
flux level was almost constant for $\sim$15 days before our
observation epoch, and thus we can reasonably assume that the shape of
the direct spectrum producing the observed halo component is the same
as that of the stable state.

To make the best estimate of the continuum spectra including low
energies below $\sim$3 keV, we derive the spectrum of the halo
component in the stable state, by taking into account the fact that
60\% of the scattered-in component is contained in the observed
spectrum. Here we refer to the cross section of dust scattering given
by \citet{dra03}, which is fine-tuned to reproduce the observed halo
intensity of GX 13+1 \citep{smi02} at its hydrogen column density
\citep{ued05}. For the oscillation states, we subtract this component
from the observed spectra and fit them by considering the opacity of
both dust-scattering and photo-electric absorption. We assume that the
halo component is the same as in the stable state and is constant
during the oscillation. The dust-scattered component is plotted in
Figure 3 (a) and (b). Its fraction in the direct (unscattered)
component is found to be 6\%, 7\%, and 15\% at 3 keV in the
oscillation-high, med and low state, respectively, which increases
toward lower energies approximately as $\propto E^{-2}$.

\subsection{Iron-K Features}

The spectra of the FI-XISs with rich photon statistics reveal that the
iron-K features are different between the four states. Figure~4 shows
the summed FI-XISs spectra in the 5--10 keV band from which the
estimated pile-up components (see above) are subtracted. The overlaid
model is the best-fit continuum model with its reflection component,
described below, from which both iron-K emission and absorption lines
are excluded to show their significance; the ratio of the data to the
model is plotted in the lower panel. In addition to a deep
absorption-edge at 7.11 keV due to the overabundance of iron in the
interstellar or circumstellar medium toward \grs, we detect an
emission line centered at 6.5--6.6 keV in both stable and oscillation
low states, whereas it is very weak in the oscillation high and medium
states. We also find a strong absorption line feature from He-like and
H-like iron ions at $\simeq 6.6$ and $\simeq 7.0$ keV in the
oscillation low state, which is weaker in the other states; in
particular, no significant absorption lines are detected in the stable
state.

\begin{figure*}
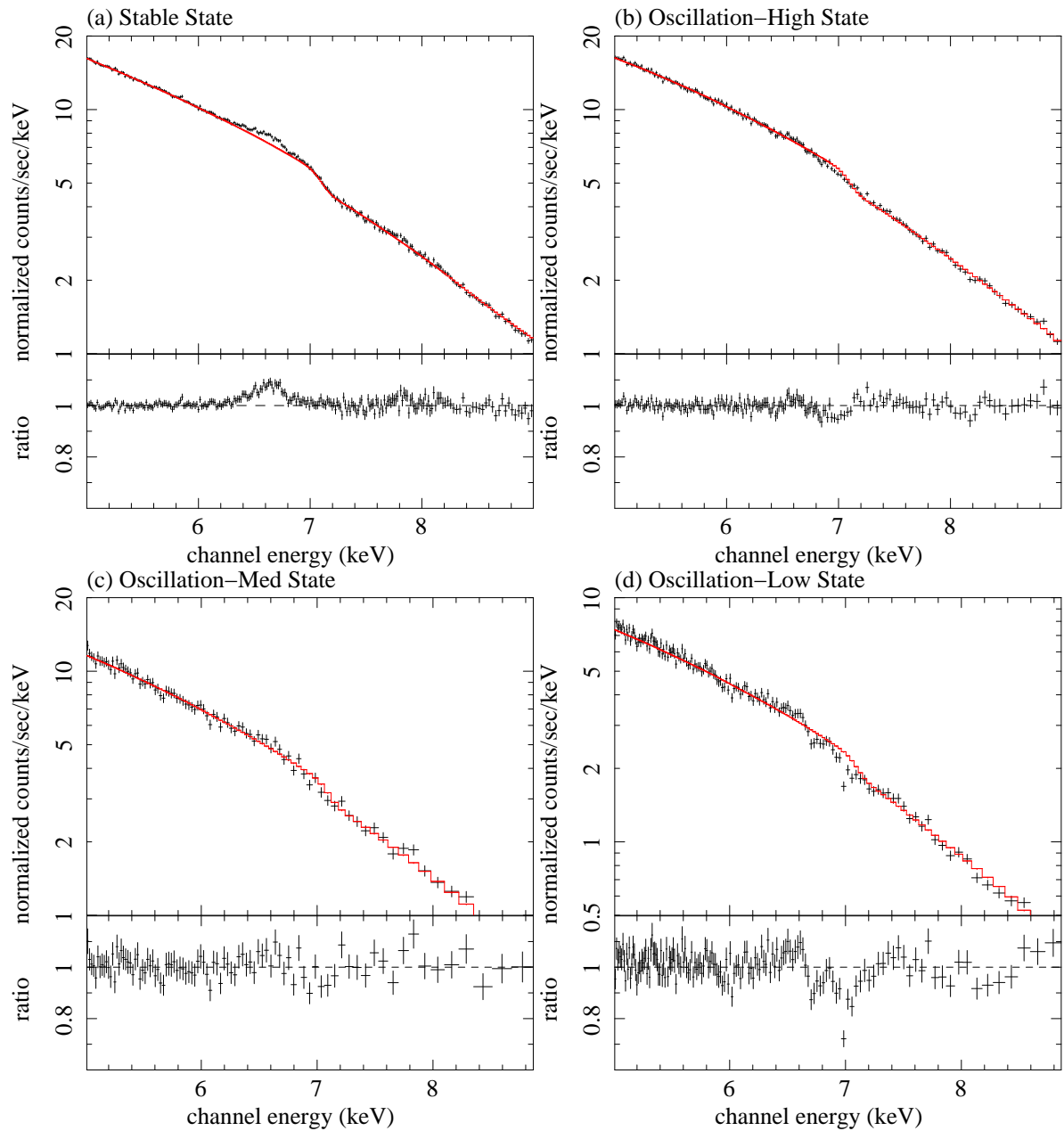

\epsscale{1.0}
\begin{center}
\includegraphics[angle=270,scale=0.5]{f4a.ps}
\includegraphics[angle=270,scale=0.5]{f4b.ps}
\includegraphics[angle=270,scale=0.5]{f4c.ps}
\includegraphics[angle=270,scale=0.5]{f4d.ps}
\end{center}
\caption{
The FI-XISs spectra in the 5--10 keV band. The solid curves (red) represent
the best-fit continuum model from which an iron-K emission and absorption 
lines are excluded. The lower panel plots the residuals in units of ratio
between the data and above model. From left to right and top to bottom,
(a) stable,
(b) oscillation-high, (c) oscillation-med, and (d) oscillation-low state.
}
\label{fig4}
\end{figure*}

We model the iron-K emission line by utilizing the {\it pexriv}
reflection code \citep{mag95}, which calculates both reflection
continuum and fluorescence line. The line profile is further blurred
with the {\it diskline} kernel \citep{fab89}, by assuming that it
originates from the accretion disk between the radii $r_{\rm in}$ and
$r_{\rm out}$ with an emissivity law of $r^{\beta}$. The free
parameters are (1) the solid angle of the reflector $\Omega$, which
basically determines the equivalent width (EW) of the emission line,
(2) ionization parameter $\xi \equiv L/n r^2$ \citep{tar69} (where
$L$, $n$, and $r$ is the luminosity, density of the reflector, and
distance from the emitter, respectively), and (3) innermost radius
$r_{\rm in}$. We fix an iron abundance at the same value as in the
interstellar absorption, an inclination of 70$^\circ$, $\beta=-3$, and
$r_{\rm out} = 10^5 r_{\rm g}$. Note that the {\it pexriv} model does
not properly calculate the additional broadening by Comptonization in
the emission line profile \citep[see][]{bal01}. Thus, we may slightly
underestimate the true equivalent width, and hence the reflection
strength in the stable state, where the ionization parameter is found
to be log $\xi$ = 2--3. In the oscillation states, the effects are
expected to be negligible due to the low ionization parameter (log
$\xi \simlt$ 2). The absorption lines are modelled by Gaussians with a
$1\sigma$ width fixed at 12 eV. The best-fit parameters are summarized
in Table~2.

\tabcolsep=2pt
\begin{deluxetable}{ccccc}
\tablenum{2}
\tablecaption{Spectral Parameters}\label{tbl-2}
\tablehead{\colhead{Parameters}
&\colhead{Stable}&\colhead{Osc-H}&\colhead{Osc-M}&\colhead{Osc-L}}
\startdata
\cutinhead{Column Densities\tablenotemark{a} ($N_{\rm H}$)}
H ($10^{22}$ cm$^{-2}$)& \multicolumn{4}{c}{2.78}\nl
Mg ($10^{22}$ cm$^{-2}$)& \multicolumn{4}{c}{6.3}\nl
Si ($10^{22}$ cm$^{-2}$)& \multicolumn{4}{c}{5.8}\nl
S ($10^{22}$ cm$^{-2}$) & \multicolumn{4}{c}{7.6}\nl
Fe ($10^{22}$ cm$^{-2}$) & \multicolumn{4}{c}{10}\nl

\cutinhead{Continuum\tablenotemark{b}}
$T_{\rm in}$ (keV) &0.53$\pm$0.04 &0.79$^{+0.02}_{-0.04}$&0.57$\pm$0.16 &0.49$^{+0.14}_{-0.19}$\nl
($T_{\rm e}$ (keV)) &2.2$^{+0.0}_{-0.2}$ &2.8$^{+0.0}_{-0.2}$ &1.7$^{+0.4}$& 1.8$^{+0.3}$ \nl
$l_{\rm h}/l_{\rm s}$&0.94$^{+0.04}_{-0.02}$ &0.84$^{+0.02}_{-0.05}$ &0.69$^{+0.10}_{-0.07}$  &0.75$^{+0.16}_{-0.07}$ \nl
$l_{\rm nt}/l_{\rm h}$&0.55$\pm0.02$&0.39$^{+0.04}_{-0.01}$&0.34$\pm$0.09  &0.17$\pm$0.05 \nl
$\tau$&8.3$\pm0.3$ &7.5$^{+0.5}_{-0.1}$ &10$_{-1.7}$ & 10$_{-1.1}$\nl
$\Gamma_{\rm inj}$ &3.7$\pm0.2$&3.1$\pm$0.3 &3.1 (fixed)& 3.1 (fixed)\nl
$R_{\rm in}^{\rm MCD}$ (km)\tablenotemark{c}&300$^{+50}_{-30}$& 120$^{+10}_{-0}$ & 150$^{+50}_{-150}$& 190$^{+160}_{-190}$ \nl
$R_{\rm in}^{\rm total}$ (km)\tablenotemark{d}&420$^{+60}_{-30}$& 210$^{+20}_{-10}$ & 360$^{+210}_{-130}$& 390$^{+380}_{-160}$ \nl
\cutinhead{Reflection\tablenotemark{e}}
$\Omega/(2\pi)$ &0.11$\pm$0.02&$<0.02$&0.10$\pm$0.07&0.54$^{+0.06}_{-0.07}$\nl
$\xi$ &530$^{+960}_{-150}$& 120 (fixed)&100$^{+360}_{-60}$&39$^{+10}_{-6}$\nl
$r_{\rm in}$ ($r_{\rm g}$) &1000$_{-200}$&290(fixed)&290(fixed)&290$^{+170}_{-140}$\nl
(E.W.)\tablenotemark{f} (eV) &33& $<$4 &13&42\nl
\cutinhead{Iron-K Absorption Features}
$E_{\rm cen}$ (keV) &&&&6.76$\pm 0.33$\nl
E.W.            (--eV) &&&&26$^{+7}_{-5}$\nl
\hline
$E_{\rm cen}$ (keV)  &6.99 (fixed)&6.99$^{+0.02}_{-0.05}$&7.01$\pm0.10$&7.02$\pm$0.02\nl
E.W.            (--eV) &$<$0.9&10.1$^{+2.7}_{-2.4}$&15$^{+8}_{-9}$&39$^{+5}_{-8}$\nl
\hline
$\tau_{\rm 8.6}^{\rm edge}$ &0 &0& 0& 0.11\nl
$\tau_{\rm 9.0}^{\rm edge}$ &0 &0.051&0.072& 0.20\nl

\cutinhead{Luminosity and Flux}
$L_{0.01-100}^{\rm MCD}$\tablenotemark{g} ($10^{38}$ erg s$^{-1}$) &9.2 &7.6 &3.0  &2.8 \nl
$L_{0.01-100}^{\rm eqpair}$\tablenotemark{g} ($10^{38}$ erg s$^{-1}$) &10.4 &15.2 &14.4 &8.8 \nl
$F_{2-10}$\tablenotemark{h} ($10^{-8}$ erg cm$^{-2}$ s$^{-1}$) &1.69 &2.88 &1.92 &1.12 \nl
$F_{10-50}$\tablenotemark{h} ($10^{-8}$ erg cm$^{-2}$ s$^{-1}$) &0.63 &1.03 &0.39 &0.17 \nl

\cutinhead{Fitting Information}
Energy Range (keV) & 2--120& 2--120& 2--50& 2--50\nl 
$\chi^2$ / d.o.f  &319/351&890/930&114/101& 127/128\nl

\enddata
\tablenotetext{a}{Equivalent hydrogen column densities as determined by \citet{ued09}, in units of Solar abundances \citep{and89} between the element and hydrogen. The abundance ratios within each group of H-He-C-N-O, Ne-Na-Mg-Al, S-Cl-Ar-Ca, and Cr-Fe-Co-Ni are fixed at the Solar values.}
\tablenotetext{b}{Fit with {\it eqpair} with a MCD spectrum as seed photons (see text). The compactness parameter for the seed photons is assumed to be $l_{\rm s}=1000$. Non-thermal electrons are considered between $\gamma_{\rm min}=1.3$ and $\gamma_{\rm max}=1000$ with a power law index of $\Gamma_{\rm inj}$.  }
\tablenotetext{c}{The normalization of the directly observed MCD component in terms of an innermost radius. The distance $D=12.5$ kpc and inclination $i=70^\circ$ are assumed without corrections for the color and boundary condition.}
\tablenotetext{d}{The innermost radius estimated from both Compton-scattered and unscattered MCD photons (see text).}
\tablenotetext{e}{Compton reflection based on the code by \citep{mag95}, blurred by the {\it diskline} kernel with an emissivity law of $r^\beta$ between the inner radius $r_{\rm in}$ and outer radius $r_{\rm out}$, where $\beta=-3$ $r_{\rm out} = 10^5 r_{\rm g}$. The accompanied iron-K emission is included.}
\tablenotetext{f}{The equivalent width of the iron-K$\alpha$ emission line
with respect to the total continuum.}
\tablenotetext{g}{Intrinsic luminosity of the direct continuum (without the reflection component) in the 0.1--100 keV band, corrected for absorption. A spherical geometry is assumed for the eqpair component, while a disk geometry is for the MCD component.}
\tablenotetext{h}{Observed flux in the 2--10 keV or 10--50 keV band.}
\tablecomments{The errors are 90\% confidence level for a single parameter.}
\end{deluxetable}

\subsection{Continuum Model}

The overall shape of the continuum in the four states is characterized
by a steep power law with a photon index of $\approx 3$ at $\sim$5--50
keV. To find an appropriate model to describe the continuum, we begin
with fitting the XIS + HXD spectra in the 2--120 keV in the stable
state, by taking into account the interstellar absorption and
dust-scattering described in \S~4.2, and the reflection component
whose parameters are determined in the previous subsection. In
addition, to include the opacity of the highly ionized gas responsible
for the iron-K absorption lines, we introduce absorption edges at
$\approx$8.6 keV and $\approx$9.0 keV corresponding to those by
\ion{Fe}{25} and \ion{Fe}{26} ions, respectively. We fix the optical
depth according to the equivalent width of the corresponding
absorption line (see Table~2), based on the results by
\citet{ued09}. We first apply a canonical model consisting of a MCD
component and a power law, which turns out to not acceptable
($\chi^2$/dof = 559/354); moreover, the best-fit results are quite
unphysical because (1) the innermost temperature of the MCD model is
high ($\sim$3 keV) with an extreme small radius ($\sim$4 km) and (2)
the power law component dominates the flux over the entire band.  This
confirms that the spectral state is very different from the canonical
high/soft state of black hole binaries. Similar fitting results are
reported by \citet{mun99} from \rxte /PCA data of \grs\ when the
0.5--10 Hz QPOs are observed.

Next, we apply a thermal Comptonization model by \citet{zyc99} with
seed photons from a MCD component, as adopted by \citet{don04},
\citet{ued09}, and \citet{vie09}. By considering that a part of the
MCD photons can escape without being Comptonized, the model is
expressed as {\it thComp + diskbb} in the XSPEC terminology. The
innermost temperature of seed photons input to the {\it thComp} model
and that of the MCD component is tied each other. We find that, while
it gives a reasonable fit to the data below $\sim$50 keV, a
significant excess remains above this energy, indicating the presence
of a non-thermal tail as discovered with OSSE \citep{zdz01} and {\it
INTEGRAL} \citep{rod08b}. Accordingly, by adding a power law component
(and its reflection) to the above model, we obtain an acceptable fit
($\chi^2$/dof = 344/350) from the broad band spectra in the stable
state. This result indicates that the spectrum can basically be
represented with Comptonization of disk photons both by thermal and
non-thermal electrons. The power-law description of non-thermal
Comptonization is unphysical, however, since it must breaks at
energies of seed photons in reality.

As a more physically self-consistent model of Comptonization, we
finally adopt the {\it eqpair} model developed by \citet{cop99},
instead of {\it thComp}. It was adopted by \citet{zdz01} and
\citet{zdz05} to fit the \rxte\ and OSSE spectra of \grs\ in
Classes~$\chi$, $\gamma$, and $\omega$ successfully. This model
computes Comptonization from a hybrid plasma consisting of thermal and
non-thermal electrons, where the physical processes are
self-consistently solved with an energy input of accelerated electrons
in a background thermal plasma. The model parameters are (1) the
optical depth for scattering $\tau$, (2) compactness parameters,
defined as $l_{\rm h,s} \equiv L_{\rm h,s} \sigma_{\rm T}/(R m_{\rm e}
c^3)$, where $L_{\rm h,s}$ is a power supplied to the plasma and that
in seed photons, $\sigma_{\rm T}$ Thomson cross section, $R$ the
plasma size, $m_{\rm e}$ the electron mass, and (3) its ratio between
non-thermal and thermal electrons, $l_{\rm nt}/l_{\rm t}$, where
$l_{\rm h} = l_{\rm nt} + l_{\rm t}$. The injected non-thermal
electrons have a power law energy distribution between the Lorentz
factors $\gamma_{\rm min}$ and $\gamma_{\rm max}$ with an index of
$\Gamma_{\rm inj}$. In the spectral fitting, we fix $l_{\rm s}=1000$,
$\gamma_{\rm min}=1.3$, $\gamma_{\rm max}=1000$.  In this model, the
energy balance between Compton and Coulomb interactions determines the
temperature of thermal electrons, which is automatically calculated
from the given parameters.

We find the {\it eqpair + diskbb} model, with the reflection of the
former component, provides an excellent description of the continuum
in all four states. The direct MCD component is required in the stable
and oscillation-high states with the F-test probability of
$7\times10^{-3}$ and $3\times10^{-8}$, respectively. Although it is
not significant in the oscillation-med and low states, we include this
component for consistency. Figure~5 plot the unfolded spectra in units
of $E I(E)$, where $I(E)$ is the energy flux. The Comptonized
component, MCD model, and the reflection component including the
iron-K emission line, are separately plotted. The best-fit continuum
parameters are summarized in Table~2, together with the observed 2--10
keV and 10--50 keV fluxes and absorption-corrected 0.01--100 keV
luminosities for the Comptonization and MCD component, where a
spherical and disk geometry is assumed, respectively.  For the
oscillation med and low states, we fix $\Gamma_{\rm inj}=3.1$, the
best-fit value obtained in the oscillation high state, since it is
difficult to constrain the slope without the coverage above 50 keV in
these states.

\begin{figure*}
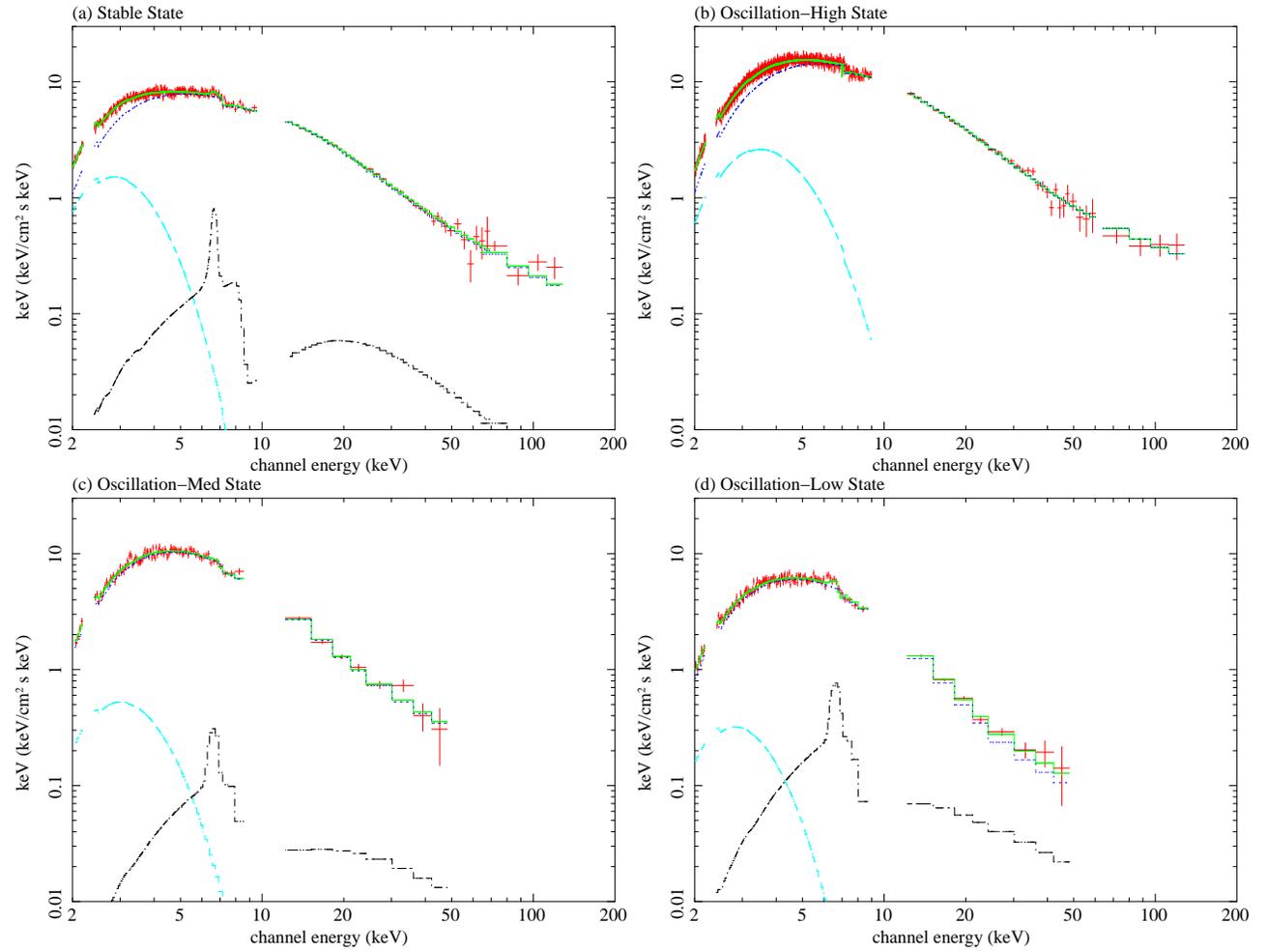

\epsscale{1.0}
\begin{center}
\includegraphics[angle=270,scale=0.35]{f5a.ps}
\includegraphics[angle=270,scale=0.35]{f5b.ps}
\includegraphics[angle=270,scale=0.35]{f5c.ps}
\includegraphics[angle=270,scale=0.35]{f5d.ps}
\end{center}
\caption{
The unfolded spectra in units of $E I(E)$ obtained from the XIS-1 and
HXD data (red, crosses) plotted with seperate contribution of the
model; from upper to lower, the total (solid, green), Comptonized
component (dot, blue), direct MCD component (dash, cyan), and
reflection component (dot-dash, black). From left to right and top to
bottom, (a) stable, (b) oscillation-high, (c) oscillation-med, and (d)
oscillation-low state.
}
\label{fig5}
\end{figure*}

\section{Discussion}

\subsection{Results Summary}

With \suzaku, we obtained high quality broad band X-ray spectra of
\grs\ over the 2--120 keV band both in the stable (Class~$\chi$) and
oscillation states (Class~$\theta$), covered with CCD energy
resolution (FWHM $\sim$120 eV at 6 keV) below 10 keV. The best
estimated interstellar absorption and dust scattering effects are
taken into account in the spectral fitting, which have been ignored in
most of previous studies of \grs. In particular, we caution that
without correct modeling of the absorption with over iron abundances,
one can easily reach wrong conclusions on the iron-K emission line
profile that requires very careful analysis of the continuum
shape. The best-fit models in units of $E I(E)$ {\it corrected for the
interstellar absorption and scattering} are plotted in Figure~6 for
the four states. Below, we interpret our results based on the iron-K
features and continuum model. We find that the iron-K features in the
oscillation low state are very similar to those in the ``soft state''
(steady State~A) of \grs\ detected by \citet{ued09} using \chandra
/HETGS. This may be expected, since our oscillation low state mostly
corresponds to State~A, though not in a steady state.

\begin{figure}
\epsscale{1.0}
\begin{center}
\includegraphics[angle=270,scale=0.35]{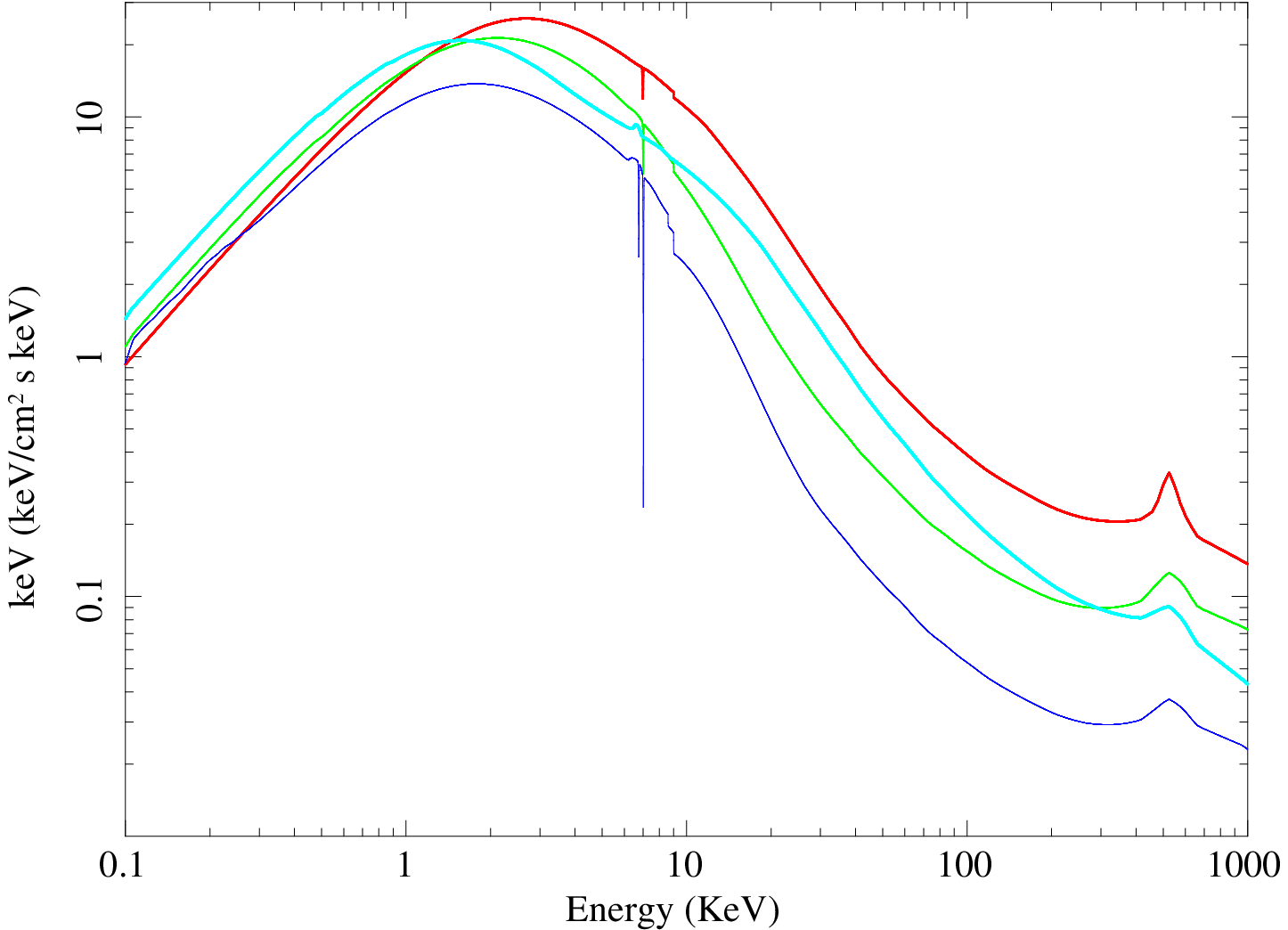}
\end{center}
\caption{
The best-fit models in units of $E I(E)$ corrected for the
interstellar absorption and dust-scattering in the four states. From
upper to lower (at 20 keV), oscillation-high (red), stable (cyan),
oscillation-med (green), and oscillation-low state (blue).
}
\label{fig6}
\end{figure}

\subsection{Change of Reflection}

We detect significant iron-K emission lines, direct evidence for the
reflection, in the stable, oscillation-med, and oscillation-low
states. The line profiles indicate that the reflector is moderately
ionized (log $\xi \sim 1-3$) and is located far away from the black
hole as constrained by its ``narrow'' line width; it is most probably
the outer parts of the accretion disk, whose inner radius is estimated
to be $\simgt 300 r_{\rm g}$ by assuming an emissivity law of
$r^{-3}$. This does not mean that the disk must be truncated at that
location. If the scale height of the central source is smaller
compared with that of the outer disk, then the emissivity law is
better expressed with an index lower than 3, which would yield a much
smaller inner radius. The absence of a relativistic broad iron-K line
is consistent with the picture that the innermost part of the disk is
covered by an optically thick hot plasma, smearing out any emission
features due to strong Comptonization.

During the limit-cycle oscillation, the EW of the iron-K emission line
(see Table~2) is the largest in the oscillation low state. It
decreases with X-ray flux, and the line {\it disappears} in the
oscillation high state. The change of the ionization parameter,
determined by the X-ray luminosity, is expected to be within a factor
of 3 during the oscillation. Hence, we must attribute this change
mainly to the variability in the solid angle of the reflector as seen
by the central source. In the stable state, the solid angle is small
$\Omega/2\pi \approx 0.1$ as well, similar to that found in the
oscillation med state.

We interpret that this is the first observational evidence for
``self-shielding'' effects of the inner part of the accretion
disk. The dynamical structure of outer parts of the accretion disk, $r
\simgt 300 r_{\rm g}$, cannot be changed on a time scale of the
limit-cycle, $\sim$1000 sec, much shorter than that of viscosity
\citep{sha73}. Therefore, the only possible mechanism that can
dramatically reduce the observed reflection strength in the
oscillation-high state must be related to the geometry in the central
source; most of the continuum X-ray emission seen by an observer
(i.e., with $i=70^\circ$) does not reach the outer parts of the disk
in this state. As we describe later, the inner accretion disk is
covered by hot corona responsible for Comptonization. Our observations
imply that the inner part of the disk becomes geometrically thick in
the oscillation high state, and Comptonized photons are shielded by
the surrounding cool region of the expanded disk when viewed at a very
high inclination angle from the outer disk. The evolution of the disk
geometry is in accordance with theoretical prediction of the slim disk
\citep{abr88} and numerical simulation \citep[e.g.,][]{ohs06}.

\subsection{Disk Wind}

As have been revealed by previous studies
\citep{kot00,lee02,ued09,nei09}, absorption lines of highly ionized
ions indicate the presence of a disk wind in the line-of-sight between
the continuum emitter and the observer. The disk wind has a velocity
of $v_{\rm wind} \approx 500$ km $s^{-1}$ and is most likely launched
at outer parts of the disk, $r_{\rm wind} \sim 10^5 r_{\rm g}$
\citep{ued09}, by irradiation from the central source. During the
limit cycle, the absorption lines are the deepest in the
oscillation-low state and become weaker as the hard X-ray flux
increases. This strongly supports the photoionization of the wind; the
corresponding ionization parameters estimated from the population
ratio between \ion{Fe}{26} and \ion{Fe}{25} are consistent with the
change of the luminosity (based on an XSTAR simulation;
\citealt{kal03}). The self-shielding is not relevant for
photo-ionization, as the wind we observe is located 20$^\circ$ above
the disk plane. The wind is most probably developed by the strong
X-ray irradiation in the oscillation-low state, as indicated by its
strong iron-K emission line. Once a disk wind is launched, it will
travel over a time scale of $r_{\rm wind}/v_{\rm wind} \sim 10^4$ sec,
and hence we can approximate that the same disk wind steadily exists
during the limit cycles.

In the stable state, we do not detect any iron-K absorption
lines. This suggests three possibilities; (1) the wind is not
developed, (2) the wind is fully ionized, or (3) the scale height of
the disk wind is too low to be observed. We consider the first
possibility unlikely because the absorption lines have been detected
in the same Class~$\chi$ state even when the hard X-ray flux is lower
than in our observation \citep{lee02}. Indeed, the presence of iron-K
emission line suggests that illumination of the outer parts of the
disk does occur, which can trigger the launch of the wind. The second
possibility cannot be ruled out, if the density of the wind is much
smaller than that develops in the oscillation states. Then, the
ionization parameter may become sufficiently large to be fully
ionized, even if the luminosity is smaller than that in the
oscillation-high state. The third possibility is also plausible.

\subsection{Origin of the Continuum}

We have shown that the broad band spectra in all the four states can
be commonly explained by Comptonization of seed disk photons off
optically-thick ($\tau \approx 7-10$), low temperature ($T_{\rm
e}\approx$2--3 keV), non-thermal hybrid plasmas. Similar results are
obtained by \citet{zdz01} in another state of \grs\ (Class~$\gamma$ or
State~B). These findings indicate that the optically thick and low
temperature Comptonization are common features in accretion flows at
high Eddington fractions, as suggested by \citet{don04} and
\citet{ued09}. In the stable and oscillation-high states, we find that
the fractional power supplied to non-thermal electrons is quite large
($l_{\rm nt}/l_{\rm h} = 0.4-0.6$) and the slope of the electron
energy distribution is steep ($\Gamma_{\rm inj} \approx$3--4) compared
with the results reported from other data of \grs\ \citep{zdz05}. This
may be related to the high luminosities in those states
($2\times10^{39}$ erg s$^{-1}$). Figure~7 plots the
absorption-corrected $E I(E)$ spectrum in the stable state with the
separate contribution from the MCD, reflection, and purely ``thermal''
Comptonization component that would be observed without non-thermal
electrons. This illustrates the importance of non-thermal processes in
these data.

\begin{figure}
\epsscale{1.0}
\begin{center}
\includegraphics[angle=270,scale=0.35]{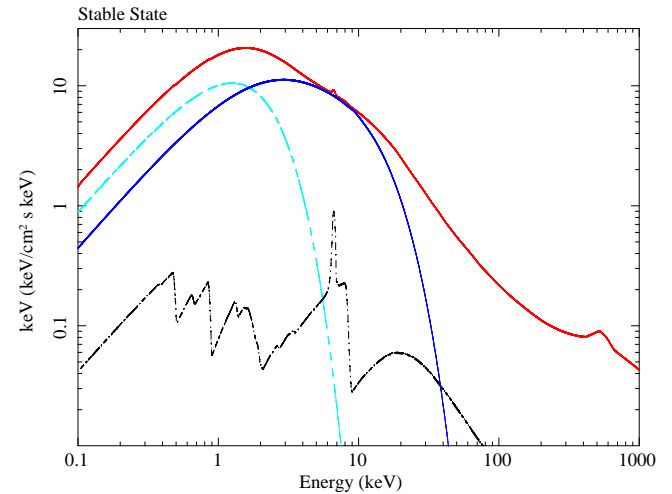}
\end{center}
\caption{
The $E I(E)$ spectrum in the stable state corrected for the
interstellar absorption and dust-scattering. The thick solid curve
(red) represents the total, the dashed curve (cyan) the direct MCD
component, the dot-dashed curve (black) the reflection component, and
the thin solid curve (blue) thermal Comptonization component that
would be observed without non-thermal electrons.
}
\label{fig7}
\end{figure}

With the {\it eqpair} modeling, we find that a direct MCD component is
necessary in addition to the Comptonized emission in the stable and
oscillation-high states, assuming the same temperature for the seed
photons and MCD component. This implies that (Case I) the Comptonizing
plasmas are patchy over the disk, or (Case II) they are located inside
a ``truncation radius'' of the standard disk. In reality, such
truncation would be smooth; the disk may gradually change into the
``disk+corona'' region, as suggested for the very high state of
canonical black hole binaries \citep[e.g.,][]{don06}. In Case II,
the assumption of the same spectrum for the direct MCD and seed
photons do not hold, and a disk radius derived from such a simple
model may be largely overestimated from the true disk size if the
energetics coupling between the disk and corona is taken into
account\citep{don06}.

From the observed flux and innermost temperature obtained from the
fit, we can estimate the innermost radius of the accretion disk both
for the direct MCD component ($R_{\rm in}^{\rm MCD}$) and that of the
seed photons for the Comptonized component ($R_{\rm in}^{\rm
Comp}$). We calculate $R_{\rm in}^{\rm Comp}$ by assuming the
conservation of photon numbers in the Comptonization process,
according to the formula (A1) of \citet{kub04} with the left-hand term
increased by a factor of two for the optically thick case. The
resultant values of $R_{\rm in}^{\rm MCD}$ and $R_{\rm in}^{\rm total}
= \sqrt{ (R_{\rm in}^{\rm MCD})^2+(R_{\rm in}^{\rm Comp})^2}$ are
listed in Table~2, without any corrections for the boundary condition
and color/effective temperature, which would increase the radius by a
factor of $\approx$1.2 \citep{kub98}. In Case I, $R_{\rm in}^{\rm
total}$ corresponds to the true innermost radius corrected for the
partial coverage by Comptonizing clouds.

We find evidence for the evolution of $R_{\rm in}^{\rm total}$ during
the oscillation, which increases from $\approx$210 km in the
oscillation-high state to $\sim$400 km in the oscillation-med/low
states, although the significance of the change is marginal due to the
large uncertainties in the latter states. They correspond to $\approx
10 r_{\rm g}$ and $\sim 20 r_{\rm g}$, respectively, by assuming the
black hole mass of 14 \solarmass. This implies that an inner disk
forms as accumulated matter rapidly accretes onto the black hole in
the flare phase. Such variation of the disk radius during limit-cycle
oscillations has been reported based on simple MCD + power law fitting
\citep[e.g.,][]{bel97b,mig03}, although we obtain larger radii than
these studies that refer only to the MCD component by ignoring
Comptonized photons. In the stable state, the radius ($R_{\rm in}^{\rm
total} \approx$420 km) is significantly larger than that in the
oscillation-high state. Thus, if Case~I is true, our results suggest
that the standard disk does not extend down to the innermost stable
circular orbit (ISCO) regardless of the black hole spin in all our
states, unlike in some cases of the ``soft state'' of \grs\
\citep{mcc06,mid06,vie09}. Such low-temperature/high-luminosity disk
emission that leads to an apparently large disk radius is observed in
the very high state of other black hole binaries \citep{don06}, when
the Eddington ratio is higher than that in the high/soft state. At
even higher Eddington fractions that trigger limit cycles, as in our
case, the situation may be the same as in the canonical very high
state.

The Comptonization parameters also show significant change during the
limit-cycle oscillation, besides the disk radius (and disk
temperature) as discussed above. In the oscillation high state, the
fractional power supplied to non-thermal electrons becomes the largest
($l_{\rm nt}/l_{\rm h} \approx 0.4$) and the electron temperature
highest ($T_{\rm e}=2.8$ keV), which decreases to $l_{\rm nt}/l_{\rm
h} \approx 0.2$ and $T_{\rm e}=1.8$ keV in the oscillation-low state,
respectively. Thus, in the flare phase, very efficient particle
acceleration must occur by converting the gravitational energy of
accreting mass. This may increase the temperature of thermal electrons
and expand the size of the corona. As the corona becomes
geometrically thick, the ``self-shielding'' effect starts to work,
preventing the Comptonized photons from irradiating the outer part of
the accretion disk (i.e., little reflection signals). A similar
situation (geometrically thick corona) may be realized in the stable
state as well, where the reflection strength is weaker compared with
the oscillation low states.

To summarize, with the simple spectra model presented above, we show
that the dramatic change of X-ray flux and spectra during the
limit-cycle oscillation can be explained mainly by (1) the
evolution of the disk geometry and (2) that of physical parameters of
Comptonizing corona, particularly the fractional power supplied to
non-thermal electrons. We note, however, that the assumption of the
MCD model (i.e., standard disk) in all states and that of the same
temperature for the seed photons of Comptonization would be 
oversimplification. For instance, disk instability theories predict
that a slim disk appears in the oscillation high state \citep{abr88},
which predicts different spectra from the MCD model. This is not ruled
out from the current simple analysis. Nevertheless, we stress that the
role of Comptonization becomes very important in determining the whole
dynamics and energy spectra of black hole accretion disks at high mass
accretion rates, which must be taken into account in any theoretical
studies.

\section{Conclusions}

\begin{enumerate}

\item We find that the \suzaku\ broad band spectra of \grs\ band both
in the stable (Class~$\chi$) and oscillation states (Class~$\theta$)
are commonly represented with a model consisting of a MCD model and
its Comptonization with a reflection component, over which the opacity
of the disk wind (including iron-K absorption lines) is applied. The
Comptonized component dominates the flux above $\sim$3 keV.

\item The Comptonization is made by optically-thick ($\tau \approx$
7--10), non-thermal / thermal ($T \approx$ 2--3 keV) hybrid
plasmas. The non-thermal electrons are produced with 10--60\% of the
total energy input to the plasma with a power law index of
$\approx$3--4, which account for the hard X-ray tail above $\sim$50
keV.

\item During the limit-cycle oscillation, the reflection strength,
estimated by the iron-K emission line, is the largest during the dip
phase but disappears in the flare phase. We interpret this as evidence
for self-shielding effects that Comptonized photons are obscured by
the surrounding cool region of the expanded disk when viewed at a very
high inclination angle from the outer disk. The evolution of the disk
geometry is in accordance with theoretical predictions.

\item The disk wind, traced by iron-K absorption lines of \ion{Fe}{25}
and \ion{Fe}{26}, always exists during the oscillation and its
ionization well correlates with the X-ray flux. This supports the
photoionization origin. In the stable state, the iron-K absorption
lines are not detected probably because it is highly ionized and/or
the scale height is small.

\item The disk parameters suggest that the inner disk structure is
similar to that in the very high state of black hole binaries. The
spectral variability in the oscillation state is explained by the
change of the disk geometry and of physical parameters of Comptonizing
corona, particularly the fractional power supplied to the
acceleration of non-thermal particles.

\end{enumerate}

\acknowledgments

We thank the \suzaku\ team and 2005 October multiwavelength campaign
team, for enabling us to perform these observations as a \suzaku\ SWG
program. Part of this work was financially supported by Grants-in-Aid
for Scientific Research 20540230, and by the grant-in-aid for the
Global COE Program ``The Next Generation of Physics, Spun from
Universality and Emergence'' from the Ministry of Education, Culture,
Sports, Science and Technology (MEXT) of Japan.


\end{document}